\documentclass[11pt,a4paper]{article}
\usepackage[hyperref]{acl2020}
\usepackage{times}
\usepackage{latexsym}
\usepackage{amsmath,amsfonts,graphicx}

\usepackage{microtype}

\aclfinalcopy 

\title{Multiresolution and Multimodal Speech Recognition with  Transformers}

\author{Georgios Paraskevopoulos \quad  Srinivas Parthasarathy \quad Aparna Khare \quad Shiva Sundaram\\
  Amazon Lab126 \\
  {\tt geopar@central.ntua.gr, \tt \{parsrini,apkhare,sssundar\}@amazon.com} \\}

\date{}

\begin{document}
\maketitle

\begin{abstract}
  This paper presents an audio visual automatic speech recognition (AV-ASR) system using a Transformer-based architecture. We particularly focus on the scene context provided by the visual information, to ground the ASR. We extract representations for audio features in the encoder layers of the transformer and fuse video features using an additional crossmodal multihead attention layer. Additionally, we incorporate a multitask training criterion for multiresolution ASR, where we train the model to generate both character and subword level transcriptions. 
  Experimental results on the How2 dataset, indicate that multiresolution training can speed up convergence by around 50\% and relatively improves  word error rate (WER) performance by upto 18\% over subword prediction models. Further, incorporating visual information improves performance with relative gains upto 3.76\% over audio only models.
  Our results are comparable to state-of-the-art Listen, Attend and Spell-based architectures.
\end{abstract}
\section{Introduction}
\label{sec:intro}
Automatic speech recognition is a fundamental technology used on a daily basis by millions of end-users and businesses.
Applications include automated phone systems, video captioning and voice assistants providing an intuitive and seemless interface between users and end systems.
Current ASR approaches rely solely on utilizing audio input to produce transcriptions.
However, the wide availability of cameras in smartphones and home devices acts as motivation to build AV-ASR models that rely on and benefit from multimodal input.

Traditional AV-ASR systems focus on tracking the user's facial movements and performing lipreading to augment the auditory inputs \cite{Potamianos97speakerindependent,mroueh2015deep,Tao2018AligningAF}.
The applicability of such models in real world environments is limited, due to the need for accurate audio-video alignment and careful camera placement.
Instead, we focus on using video to contextualize the auditory input and perform multimodal grounding. For example, a basketball court is more likely to include the term ``lay-up'' whereas an office place is more likely include the term ``lay-off''.
This approach can boost ASR performance, while the requirements for video input are kept relaxed \cite{how2-baseline, av-grounding-asr}. Additionally we consider a multiresolution loss that takes into account transcriptions at the character and subword level. We show that this scheme regularizes our model showing significant improvements over subword models. Multitask learning on multiple levels has been previously explored in the literature, mainly in the context of CTC~\cite{sanabria2018hierarchical,krishna2018hierarchical,ueno2018acoustic}. A mix of seq2seq and CTC approaches combine word and character level~\cite{kremer2018inductive,ueno2018acoustic} or utilize explicit phonetic information \cite{toshniwal2017multitask,sanabria2018hierarchical}.



Modern ASR systems rely on end-to-end, alignment free neural architectures, i.e. CTC  \cite{ctc} or sequence to sequence models \cite{rnnt,deep-cnn}.
The use of attention mechanisms significantly improve results in \cite{attention-asr} and \cite{las}.
Recently, the success of transformer architectures for NLP tasks \cite{transformer,bert,transformer-xl} has motivated speech researchers to investigate their efficacy in end-to-end ASR \cite{karita2019comparative}.
Zhou et. al., apply an end-to-end transformer architecture for Mandarin Chinese ASR \cite{zhou2018syllable}.
Speech-Transformer extends the scaled dot-product attention mechanism to $2$D and achieves competitive results for character level recognition  \cite{speech-transformer, improving-speech-transformer}.
Pham et. al. introduce the idea of stochastically deactivating layers during training to achieve a very deep model \cite{very-deep-transformer}.
A major challenge of the transformer architecture is the quadratic memory complexity as a function of the input sequence length.
Most architectures employ consecutive feature stacking \cite{very-deep-transformer} or CNN preprocessing \cite{speech-transformer,karita2019comparative} to downsample input feature vectors.
\citet{vggtransformer} use a VGG-based input network to downsample the input sequence and achieve learnable positional embeddings.

\begin{figure*}[htb]
	\centering
	\includegraphics[width=.8\linewidth]{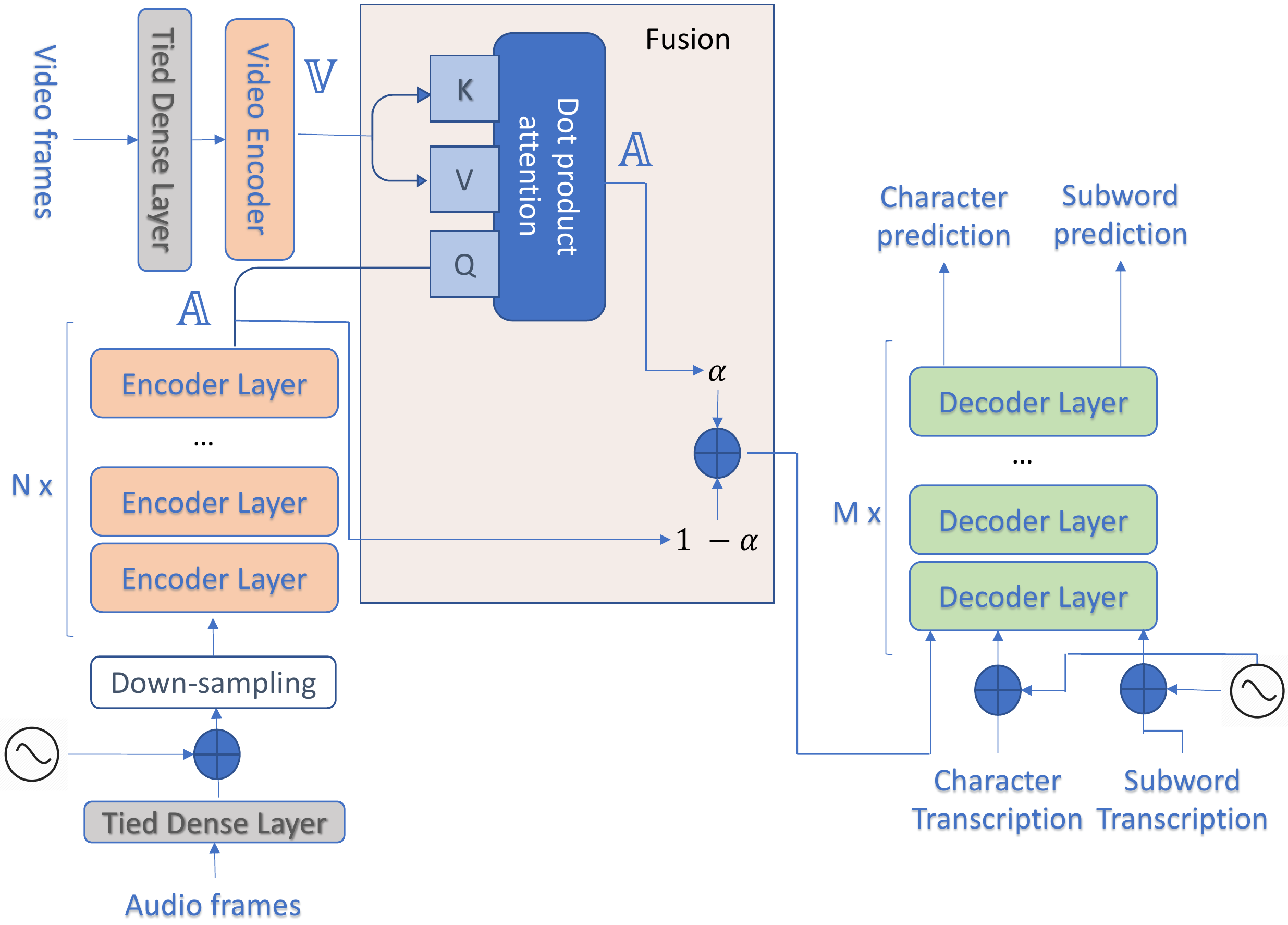}
	\caption{Overall system architecture. A cross-modal scaled dot-product attention layer is used to project the visual data into the audio feature space followed by an additive fusion.}
	\label{fig:architecture}
\end{figure*}

Multimodal grounding for ASR systems has been explored in \cite{how2-baseline}, where a pretrained RNN-based ASR model is finetuned with visual information through Visual Adaptive Training.
Furthermore, \citet{av-grounding-asr} use a weakly supervised semantic alignment criterion to improve ASR results when visual information is present.
Multimodal extensions of the transformer architecture have also been explored. These extensions mainly fuse visual and language modalities in the fields of Multimodal Translation and Image Captioning.
Most approaches focus on using the scaled dot-product attention layer for multimodal fusion and cross-modal mapping.
\citet{afouras2018deep} present a transformer model for AV-ASR targeted for lip-reading in the wild tasks. It uses a self attention block to encode the audio and visual dimension independently. A decoder individually attends to the audio and video modalities producing character transcriptions. In comparison our study uses the video features to provide contextual information to our ASR.
\citet{libovicky-etal-2018-input} employ two encoder networks for the textual and visual modalities and propose four methods of using the decoder attention layer for multimodal fusion, with hierarchical fusion yielding the best results.
\citet{yu2019multimodal} propose an encoder variant to fuse deep, multi-view image features and use them to produce image captions in the decoder.
\citet{le-etal-2019-multimodal} use cascaded multimodal attention layers to fuse visual information and dialog history for a multimodal dialogue system.
\citet{cross-modal-transformer} present Multimodal Transformers, relying on  a deep pairwise cascade of cross-modal attention mechanisms to map between modalities for multimodal sentiment analysis.

In relation to the previous studies, the main contributions of this study are a) a fusion mechanism for audio and visual modalities based on the crossmodal scaled-dot product attention, b) an end to end training procedure for multimodal grounding in ASR and c) the use of a multiresolution training scheme for character and subword level recognition in a seq2seq setting without relying on explicit phonetic information.
We evaluate our system in the $300$ hour subset of the How2 database \cite{sanabria18how2}, achieving relative gains up to 3.76\% with the addition of visual information. Further we show relative gains of 18\% with the multiresolution loss. Our results are comparable to state-of-the-art ASR performance on this database.

\section{Proposed Method}
Our transformer architecture uses two transformer encoders to individually process acoustic and visual information (Fig. \ref{fig:architecture}).
Audio frames are fed to the first set of encoder layers.
We denote the space of the encoded audio features as the audio space $\mathbb{A}$.
Similarly, video features are projected to the video space $\mathbb{V}$ using the second encoder network.
Features from audio and visual space are passed through a tied feed forward layer that projects them into a common space before passing them to their individual encoder layers respectively. This tied embedding layer is important for fusion as it helps align the semantic audio and video spaces.
We then use a cross-modal attention layer that maps projected video representations to the projected audio space (Section \ref{ssec:cm_attention}).
The outputs of this layer are added to the original audio features using a learnable parameter $\alpha$ to weigh their contributions.
The fused features are then fed into the decoder stack followed by dense layers to generate character and subword outputs. For multiresolution predictions (Section \ref{ssec:multiresolution}), we use a common decoder for both character and subword level predictions, followed by a dense output layer for each prediction. This reduces the model parameters and enhances the regularization effect of multitask learning.

\subsection{Cross-modal Attention}
\label{ssec:cm_attention}
Scaled dot-product attention operates by constructing three matrices, $K$, $V$ and $Q$ from sequences of inputs.
$K$ and $V$ may be considered keys and values in a ``soft'' dictionary, while $Q$ is a query that contextualizes the attention weights.
The attention mechanism is described in Eq.~\ref{eq:att}, where $\sigma$ denotes the softmax operation.
\begin{equation}
    Y = \sigma(KQ^T)V
\label{eq:att}
\end{equation}

The case where $K$, $V$ and $Q$ are constructed using the same input sequence consists a self-attention mechanism.
We are interested in cross-modal attention, where $K$ and $V$ are constructed using inputs from one modality $\mathbb{M}_1$, video in our case (Fig.~\ref{fig:architecture}) and $Q$ using another modality $\mathbb{M}_2$, audio.
This configuration as an effective way to map features from $\mathbb{M}_1$ to $\mathbb{M}_2$ \cite{cross-modal-transformer}.
Note, that such a configuration is used in the decoder layer of the original transformer architecture \cite{transformer} where targets are attended based on the encoder outputs.
\subsection{Multiresolution training}
\label{ssec:multiresolution}
We propose the use of a multitask training scheme where the model predicts both character and subword level transcriptions.
We jointly optimize the model using the weighted sum of character and subword level loss, as in Eq.~\ref{eq:loss}:
\begin{equation}
    L = \gamma * L_{subword} + (1 - \gamma) * L_{character}
\label{eq:loss}
\end{equation}
\noindent where $\gamma$ is a hyperparameter that controls the importance of each task.

The intuition for this stems from the reasoning that character and subword level models perform different kinds of mistakes.
For character prediction, the model tends to predict words that sound phonetically similar to the ground truths, but are syntactically disjoint with the rest of the sentence.
Subword prediction, yields more syntactically correct results, but rare words tend to be broken down to more common words that sound similar but are semantically irrelevant.
For example, character level prediction may turn ``\textit{old-fashioned}'' into ``\textit{old-fashioning}'', while subword level
turns the sentence ``\textit{ukuleles are different}'' to ``\textit{you go release are different}''.
When combining the losses, subword prediction, which shows superior performance is kept as the preliminary output, while the character prediction is used as an auxiliary task for regularization.


\section{Experimental Setup}
We conduct our experiments on the How2 instructional videos database \cite{sanabria18how2}.
The dataset consists of $300$ hours of instructional videos from the YouTube platform.
These videos depict people showcasing particular skills and have high variation in video/audio quality, camera angles and duration.
The transcriptions are mined from the YouTube subtitles, which contain a mix of automatically generated and human annotated transcriptions.
Audio is encoded using $40$ mel-filterbank coefficients and $3$ pitch features with a frame size of $10$ ms, yielding $43$-dimensional feature vectors.
The final samples are segments of the original videos, obtained using word-level alignment.
We follow the video representation of the original paper \cite{how2-baseline}, where a $3$D ResNeXt-101 architecture, pretrained on action recognition, is used to extract $2048$D features \cite{hara2018can}.
Video features are average pooled over the video frames yielding a single feature vector. For our experiments, we use the train, development and test splits proposed by \cite{sanabria18how2}, which have sizes $298.2$ hours, $3.2$ hours and $3.7$ hours respectively.

Our model consists of $6$ encoder layers and $4$ decoder layers.
We use transformer dimension $480$, intermediate ReLU layer size $1920$ and $0.2$ dropout.
All attention layers have $6$ attention heads.
The model is trained using Adam optimizer with learning rate $10^{-3}$ and $8000$ warmup steps.
We employ label smoothing of $0.1$.
We weigh the multitask loss with $\gamma=0.5$ which gives the best performance.
A coarse search was performed for tuning all hyperparameters over the development set.
For character-level prediction, we extract $41$ graphemes from the transcripts.
For subword-level prediction, we train a SentencePiece tokenizer \cite{kudo2018sentencepiece} over the train set transcriptions using byte-pair encoding and vocabulary size $1200$.
For decoding we use beam search with beam size $5$ and length normalization parameter $0.7$. 
We train models for up to $200$ epochs and the model achieving the best loss is selected using early stopping. Any tuning of the original architecture is performed on the development split.
No language model or ensemble decoding is used in the output.
\section{Results and Discussion}
\begin{table}[tb]
	
	\begin{center}
		\begin{tabular}{|c c c |}
			\hline
			Input handling & Recognition level & WER  \\ [0.5ex]
			\hline\hline
			Filtering      & Character & $33.0$ \\
			\hline
			Filtering      & Subword   & $29.7$ \\
			\hline
			Chunking & Character & $31.3$  \\
			\hline
			Chunking & Subword   & $29.9$ \\
			\hline
			Stacking & Character & $28.3$  \\
			\hline
			Stacking & Subword   & $26.1$ \\
			\hline
			Stacking & MR & $\mathbf{21.3}$ \\
			\hline
			
		\end{tabular}
	\end{center}
	\caption{Results for different methods of input filtering for different prediction resolutions. \emph{MR} stands for multiresolution.}
	\label{tab:subword-character}
\end{table}

One of the challenges using scaled dot-product attention is the quadratic increase of layerwise memory complexity as a function of the input sequence length.
This issue is particularly prevalent in ASR tasks, with large input sequences. We explore three simple approaches to work around this limitation.
First, we filter out large input sequences ($x>15s$), leading to loss of $100$ hours of data.
Second we, chunk the input samples to smaller sequences, using forced-alignment with a conventional DNN-HMM model to find pauses to split the input and the transcriptions.
Finally, we stack $4$ consecutive input frames into a single feature vector, thus reducing the input length by $4$. Note that this only reshapes the input data as the dimension of our input is increased by the stacking process
\footnote{We tried to use the convolutional architecture from \cite{vggtransformer}, but it failed to converge in our experiments, possibly due to lack of data}.
Results for the downsampling techniques for character and subword level predictions are summarized in Table~\ref{tab:subword-character}.
We observe that subword-level model performs better than the character level (upto 10\% relative) in all settings.
This can be attributed to the smaller number of decoding steps needed for the subword model, where error accumulation is smaller.
Furthermore, we see that the naive filtering of large sequences yields to underperforming systems due to the large data loss. Additionally, we see that frame stacking has superior performance to chunking.
This is not surprising as splitting the input samples to smaller chunks leads to the loss of contextual information which is preserved with frame stacking.
We evaluate the proposed multiresolution training technique with the frame stacking technique, observing a significant improvement(18.3\%) in the final WER.
We thus observe that predicting finer resolutions as an auxiliary task can be used as an effective means of regularization for this sequence to sequence speech recognition task. Furthermore, we have empirically observed that when training in multiple resolutions, models can converge around 50\% faster than single resolution models.

\begin{table}[tb]
	\small
	\begin{center}
		\begin{tabular}{|c c c c|}
			\hline
			&  &   & $\Uparrow$ \\
			Features       &  Level     &   WER   & over audio \\
			[0.5ex]
			\hline\hline
			Audio & Subword & $26.1$ & -\\
			\hline
			Audio + ResNeXt & Subword &  $25.0$ & $3.45\%$ \\
			\hline
			\hline
			Audio & MR & $\mathbf{21.3}$ & - \\
			\hline
			Audio + ResNeXt & MR & $20.5$ & $3.76\%$ \\
			\hline
			\hline
			Audio (B) & Subword & $\mathbf{19.2}$ & - \\
			\hline
			Audio + ResNext (B) & Subword & $\mathbf{18.4}$ & $3.13\%$ \\
			\hline
		\end{tabular}
	\end{center}
	\caption{Comparison of audio only ASR models versus AVASR models with ResNeXt image features. \emph{MR} stands for multiresolution. \emph{(B)} shows the results for the LAS model \cite{how2-baseline}}
	\vspace{-0.6cm}
	\label{tab:multimodal}
	
\end{table}
Next, we evaluate relative performance improvement obtained from utilizing the visual features (Table~\ref{tab:multimodal}).
We observe that incorporating visual information improves ASR results. Our AV-ASR system yields gains $>3\%$ over audio only models for both subword and multiresolution predictions. 
Finally, we observe that while the Listen, Attend and Spell-based architecture of \cite{how2-baseline} is slightly stronger than the transformer model, the gains from adding visual information is consistent across models. It is important to note that our models are trained end-to-end with both audio and video features.

An important question for real-world deployment of multimodal ASR systems is their performance when the visual modality is absent.
Ideally, a robust system satisfactorily performs when the user's camera is off or in low light conditions.
We evaluate our AV-ASR systems in the absence of visual data with the following experiments - a) replace visual feature vectors by zeros b) initialize visual features with gaussian noise with standard deviation $0.2$ c) tweak the value $\alpha$ to 0 on inference, gating the visual features completely. Table~\ref{tab:audio-only} shows the results for the different experiments. Results indicate gating visual inputs works better than zeroing them out. Adding a gaussian noise performs best which again indicates the limited availability of data. Overall, in the absence of visual information, without retraining, the AV-ASR model relatively worsens by 6\% compared to audio only models. 

\begin{table}[tb]

    \begin{center}
        \begin{tabular}{|c c |}
        \hline
        Missing input handling & WER  \\ [0.5ex]
        \hline\hline
        Zeros &  23.1 \\
        \hline
        Gaussian Noise $\sigma$=0.2 & 22.6 \\
        \hline
        Gating visual input $\alpha$=0 & 22.8 \\
        \hline
       \end{tabular}
   \end{center}
	\vspace{-0.3cm}
   \caption{Experimental evaluation of AV-ASR model for handling missing visual input. Here $\sigma$ denotes the standard deviation of the noise}
   \label{tab:audio-only}
\end{table}

\section{Conclusions}
This paper explores the applicability of the transformer architecture for multimodal grounding in ASR. Our proposed framework uses a crossmodal dot-product attention to map visual features to audio feature space. Audio and visual features are then combined with a scalar additive fusion and used to predict character as well as subword transcriptions. We employ a novel multitask loss that combines the subword level and character losses. Results on the How2 database show that a) multiresolution losses regularizes our model producing significant gains in WER over character level and subword level losses individually b) Adding visual information results in relative gains of 3.76\% over audio model's results validating our model.

Due to large memory requirements of the attention mechanism, we apply aggressive preprocessing to shorten the input sequences, which may hurt model performance. In the future, we plan to alleviate this by incorporating ideas from sparse transformer variants \cite{reformer,sparsetransformer}.
Furthermore, we will experiment with more ellaborate, attention-based fusion mechanisms.
Finally, we will evaluate the multiresolution loss on larger datasets to analyze it's regularizing effects.
\bibliographystyle{acl_natbib}

\bibliography{acl2020}

\end{document}